\documentclass[dvipdfm]{llncs}
\usepackage[latin1]{inputenc}
\usepackage{url,amsfonts,epsfig}
\usepackage{amsmath,latexsym,amssymb}
\usepackage{color,caption}
\usepackage{graphicx, subfigure}
\usepackage{footmisc}
\usepackage{algorithm,algorithmic}

\newtheorem{Theorem}{Theorem}

\newtheorem{Proposition}{Proposition}
\newtheorem{Lemma}{Lemma}
\newtheorem{Definition}{Definition}

\newcommand{\dmin}{d_{min}}
\newcommand{\dmax}{d_{max}}

\title{On relaxing the constraints in pairwise compatibility graphs}
\titlerunning{On relaxing the constraints in pairwise compatibility graphs}  

\author{Tiziana Calamoneri \and Rossella Petreschi \and Blerina Sinaimeri}
\authorrunning{T. Calamoneri, R. Petreschi and B. Sinaimeri}   
\institute{Department of Computer Science \\
    ``Sapienza'' University of Rome - Italy\\
  via Salaria 113, 00198 Roma, Italy.\\
\email{e-mail: \{calamo, petreschi, sinaimeri\}@di.uniroma1.it}
}

\begin{document}

\maketitle              

\begin{abstract}
A graph $G$ is called a pairwise compatibility graph (PCG) if there exists an edge weighted tree $T$ and two non-negative real numbers $d_{min}$ and $d_{max}$ such that each leaf $l_u$ of $T$ corresponds to a vertex $u \in V$ and there is an edge $(u,v) \in E$ if and only if $d_{min} \leq d_T (l_u, l_v) \leq d_{max}$ where $d_T (l_u, l_v)$ is the sum of the weights of the edges on the unique path from $l_u$ to $l_v$ in $T$.  In this paper we analyze the class of PCG in relation with two particular subclasses resulting from the the cases where $\dmin=0$ (LPG) and $\dmax=+\infty$ (mLPG). In particular, we show that the union of LPG and mLPG does not coincide with the whole class PCG, their intersection is not empty, and that neither of the classes LPG and mLPG is contained in the other.  Finally, as the graphs we deal with belong to the more general class of split matrogenic graphs,  we focus on this class of graphs for which we try to establish the membership to the PCG class.

\end{abstract}

{\bf keywords:} PCG, leaf power graph, threshold graphs, matrogenic graphs.

\section{Introduction}

Given an edge weighted tree $T$, let $d_{min}$ and $d_{max}$ be two nonnegative real numbers with $d_{min} \leq d_{max}$. For any two leaves $l_1$ and $l_2$ of the tree $T$, we denote by $d_T (l_1, l_2)$ the sum of the weights of the edges on the unique path from $l_1$ to $l_2$ in $T$. Starting from $T$, $\dmin$ and $\dmax$, it can be easily constructed a \textsl{pairwise compatibility graph }of $T$, i.e. a graph $G(V,E)$ where each vertex $u \in V$ corresponds to a leaf $l_u$ of $T$ and there is an edge $(u,v) \in E$ if and only if $d_{min} \leq d_T (l_u, l_v) \leq d_{max}$. We will denote such a graph $G$ by $PCG(T,d_{min}, d_{max})$. Consequently, we say that a graph $G$ is a pairwise compatibility graph (PCG) if there exists an edge weighted tree $T$ and two nonnegative real numbers  $d_{min}$ and $d_{max}$ such that $G=PCG(T,d_{min}, d_{max})$. Determine whether a graph $G$ is a PCG seems in general difficult even if at the beginning, when the problem arose in a computational biology context \cite{Kal03}, it was conjectured that every graph was a PCG. Nowadays it is known that this conjecture is false \cite{YBR10}, while it is proved that some specific classes of graphs e.g., graphs with five nodes or less \cite{P02}, cliques and disjoint union of cliques \cite{B}, chordless cycles and single chord cycles \cite{YHR09} and some particular subclasses of bipartite graphs \cite{YBR10}, are PCG.

The pairwise compatibility concept is defined with respect to two bounds concerning $\dmin$ and $\dmax$. If we relax these conditions, requiring only that the distance between some pair of leaves is smaller than or equal to $\dmax$ (i.e. we set $\dmin=0$) then we are considering a particular subclass of PCG graphs, namely the \textsl{leaf power} graphs (LPG). More formally, a graph $G(V,E)$ is a leaf power if there exists a tree $T$ and a nonnegative number $\dmax$ such that there is an edge $(u,v)$ in $E$ if and only if for their corresponding leaves $l_u, l_v$ we have $d_T(l_u, l_v) \leq \dmax$ (see \cite{NRTh02}). Although there has been a lot of works on this class of graphs \cite{B}, a completely description of leaf power graphs is still unknown.

To the best of our knowledge, nothing is known in literature concerning the subclass of PCG when the constraint concerns only the minimum distance, i.e. there is an edge in $E$ if and only if the corresponding leaves are at a distances greater than $k$ in the tree (observe that in this case we set $\dmax=+ \infty$). In this paper we introduce this new concept and exploit the relations between the new defined class and the two known classes LPG and PCG.

The paper is organized as follows: in Section \ref{sec:preliminaries} we introduce some terminologies and recall some known concepts that we will use in the forthcoming work. Then, we define the new subclass of PCG, namely mLPG, characterized by the use of $d_{min}$ only. Next, in Section \ref{sec:relations} we study the relations between the classes PCG, LPG and mLPG. In particular, we show that the union of LPG and mLPG does not coincide with the whole class PCG, their intersection is not empty, and neither of the classes LPG and mLPG is contained in the other. All the graphs we furnish as examples in Section \ref{sec:relations} are particular cases of the more general class of split matrogenic graphs. Hence, in Section \ref{sec:matrogenic} we focus on the class of split matrogenic graphs trying to determine if it belongs to the PCG class. We prove that many split matrogenic graphs are PCG. However, the membership to PCG class of one particular subclass of split matrogenic graph remains an open problem that is reported in the final Section \ref{sec:conclusion} together with some other open problems.
\section{Preliminaries}\label{sec:preliminaries}

In this section we introduce some definitions and some concepts that we use in the rest of  this paper. 

When we say that a {\em tree} $T$ is {\em weighted}, we mean that it is edge weighted, that is each edge is assigned a number as its weight. In this paper we consider only weighted trees and graphs that are connected.

A {\em caterpillar} is a tree in which all the vertices are within distance one of a central path which is called the {\em spine}.

A graph $G=(K,S,E)$ is said to be\textit{ split} if there is a vertex partition \mbox{$V=K \cup S$} such that the subgraphs induced by $K$ and $S$ are complete and stable, respectively.

Given two split graphs $G_1=(K_1,S_1,E_1)$ and $G_2=(K_2,S_2,E_2)$  their \textit{composition} $G_1\circ G_2$ is formed by taking the disjoint union of $G_1$ and $G_2$ and adding all the edges  $\{ u, v\}$ such that $u \in K_1$ and $v \in V(G_2)$. Observe that $G_1\circ G_2$ is again a split graph.

A set $M$ of edges is a {\em perfect matching} of dimension $m$ of $A$ onto $B$ if and only if $A$ and $B$ are disjoint subsets of vertices of cardinality $m$ and each vertex in $A$ is adjacent to exactly one vertex in $B$ and no two edges share a point. We say that the split graph $G=(K,S,E)$ is a  {\em split matching }if the subset of edges in $E$ not belonging to the clique forms a perfect matching. We denote by $\mathcal{SM}$ the class of split matching graphs.

An \textit{antimatching} of dimension $m$ of $A$ onto $B$ is a set of edges such that the non edges between $A$ and $B$ form a perfect matching. We say that the split graph $G=(K,S,E)$ is a \textit{split antimatching }if the subset of edges in $E$ not belonging to the clique forms an antimatching. We denote by $\mathcal{SA}$ the class of split antimatching graphs.

A {\em cactus} is a connected graph in which any two simple cycles have at most one vertex in common. Equivalently, every edge in such a graph may belong to at most one cycle. We will denote by $\mathcal{C}$ the class of cacti with at least a cycle of length $n \geq 5$.

Given a connected graph $G$ whose distinct vertex degrees are \mbox{$: \delta_1 >  \ldots > \delta_r$}, we define $B_i=\{ v \in V(G): deg(v)= d_i\}$, for any $i=1, \ldots, r$.  The sets $B_i$ are usually referred as {\em boxes }and the sequence $B_1, \ldots, B_r$ is called the \textit{degree partition of} $G$ \textit{into boxes}.

Given a graph $G$ with  degree partition $B_1, \ldots, B_r$, $G$ is a  {\em threshold graph} if and only if for all $u \in B_i$, $v \in B_j$, $u \neq v$, we have $(u,v) \in E(G)$ if and only if $i+j \leq r+1$. We will denote by $\mathcal{T}$ the class of threshold graphs.

\smallskip

Now, we introduce a new subclass of PCG, namely mLPG as follows: 

\begin{Definition}
A graph $G(V,E)$ is an mLPG if there exists a tree $T$ and an integer $\dmin$ such that there is an edge $(u,v)$ in $E$ if and only if for their corresponding leaves $l_u, l_v$ in $T$ we have $d_T(l_u, l_v) \geq \dmin$. 
\end{Definition}
Note that for the sake of simplicity and homogeneity of the paper, here we slightly abuse notation as these graphs are not power of trees.

\medskip
\noindent
In what follows we will often make use of the following simple observation.

\begin{Proposition}\label{prop:technical}
Let $G$ be a  graph that does not belong to some class $L$ from $\{PCG, LPG, mLPG\}$ then every graph $H$ that contains $G$ as an induced subgraph, does not belong to $L$ either.
\end{Proposition}

Given two vertices $u,v$ in a tree $T$, we denote by $P_{uv}$ the unique path in $T$ connecting the vertices $u$ and $v$. A {\em subtree induced by a set of leaves} of $T$ is the minimal subtree of $T$ which contains those leaves. We denote by $T_{uvw}$ the subtree of a tree induced by three leaves $u, v$ and $w$. 

The following technical lemma will turn out to be very useful in the forthcoming results.

\begin{Lemma}\label{lem:technical} \cite{YBR10}
Let $T$ be an edge weighted tree, and $u, v$ and $w$ be three leaves of $T$
such that $P_{uv}$ is the largest path in $T_{uvw}$. Let $x$ be a leaf of $T$ other than $u, v$ and
$w$. Then, either $d_T(w, x) \leq d_T (u, x)$ or $d_T(w, x)\leq d_T(v, x)$.
\end{Lemma}

\begin{figure}[h]
	\centering
	\includegraphics[scale=0.4]{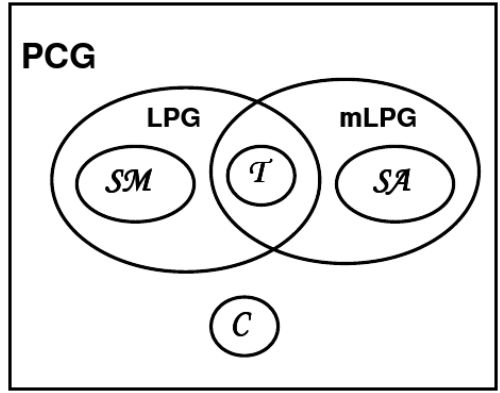}
\caption{Relationships between PCG, LPG and mLPG.}
\label{fig:classes}
\end{figure}

\section{Relationships between PCG, LPG and mLPG  }\label{sec:relations}

In this section we study the relationships between the classes of $PCG$, $LPG$ and $mLPG$. First, in Subsection \ref{subsec:union} we show that the union of $mLPG$ and $LPG$ does not contain the whole class of $PCG$. Next, in Subsection \ref{subsec:intersection} we show that their intersection $LPG \cap mLPG$ is not empty, by proving that threshold graphs belong to both classes. Finally, in Subsection \ref{subsec:proper} we show that neither of the classes $LPG$ and $mLPG$ is contained in the other one by providing for each of these classes a particular graph which is proper to it. These relations are graphically shown in Figure \ref{fig:classes}.

\subsection{$PCG \supset LPG \cup mLPG$} \label{subsec:union}
In this subsection we prove that the $PCG$ class does not coincide with the union of $LPG$ and $mLPG$. Indeed, in \cite{YHR09} it is proved that any cycle is a PCG. Now, it is well-known (see, for example, \cite{B}) that LPG is a subclass of strongly chordal graphs and clearly cycles of length $n \geq 5$ are not strongly chordal, so they are not LPG. The following lemma states that cycles do not belong to mLPG, deducing that $(LPG \cup mLPG) \subset PCG$.

\begin{Lemma}
Let $C_n$ be a cycle of length $n \geq 5$, then $C_n \not \in mLPG$.
\end{Lemma}
\proof
Let $v_1, \ldots v_n$ be the ordered vertices of a cycle $C_n$ with $n \geq 5$. Suppose by contradiction that $C_n=mLPG (T,\dmin)$ and let $l_i$
be the leaf in $T$ corresponding to the vertex $v_i$, for any $i \leq n$.  Let $v_1, v_2, v_3$ be the first three consecutive vertices in $C_n$ and consider the largest path in $T_{l_1 l_2 l_3}$. As $(v_1,v_3) \not\in E$ (as $n\geq 5 $) then $d_T(l_1,l_3) < \dmin$. Hence, the largest path must be one from $P_{l_1 l_2}$ and $P_{l_2 l_3}$. 

Suppose first the largest path is $P_{l_1 l_2}$. Using Lemma \ref{lem:technical} with $x=l_4$ we have that either $\dmin \leq d_T(l_4,l_3) \leq d_T(l_4,l_2)$ or $\dmin \leq d_T(l_4,l_3) \leq d_T(l_4,l_1)$, deducing that at least one between the $(v_4,v_2)$ and $(v_4,v_1)$ must be an edge in $C_n$, a contradiction. 

If $P_{l_2l_3}$ is the largest path, we arrive at the same result  by taking this time $x=l_n$. This concludes the proof. \qed

Easily, in view of Proposition \ref{prop:technical} the class $\mathcal{C}$ of cacti with at least one cycle of length $n \geq 5$ does not belong either to LPG or to mLPG. 

\subsection{$LPG \cap mLPG \neq \emptyset$}\label{subsec:intersection}

In this subsection we prove that the intersection of LPG and mLPG is not empty by showing that threshold graphs belong to $LPG  \cap mLPG$. 

\begin{Theorem}\label{theo:threshold}
Let $G$ be a threshold graph. Then $G \in LPG  \cap mLPG$ and it is polynomial to find the tree $T$ and the values $\dmin, \dmax$ associated to $G$.
\end{Theorem}

\proof
Let $G$ be a threshold graph on $n$ vertices and let $B_1, \ldots, B_r$ be the degree partition of $G$.  We consider an $n$-leaf star with center a vertex $c$, as the tree $T$. 

To prove that $G \in LPG$, for each vertex $v$ of $G$, assign weight  $i$ to the edge $(l_v,c)$ in $T$ if $v \in B_i$. Define $\dmax=r+1$.  As for each $u \in B_i$, $v \in B_j$, $u \neq v$, we have $(u,v) \in E(G)$ if and only if $i+j \leq r+1$ it is straightforward that $G$ is a PCG of $T$ with $\dmax$.

On the other hand, to prove $G \in mLPG$ for any $v \in V(G)$ assign $r+1-i$ to the edge $(l_v,c)$ in $T$ if $v \in B_i$.  Note that, as $i \leq r$ we assign nonnegative weights to the edges of the star. Define $\dmin=r+1$. For any two vertices $v \in B_i$ and $u \in B_j$, we have that if $i + j \leq r+1$ (meaning that $(u,v) \in E(G)$) then $d_T(l_u,l_v)=2(r+1)-(i+j) \geq r+1=\dmin$. Otherwise, if $i + j > r+1$ (meaning that $(u,v) \not\in E(G)$) then $d_T(l_u,l_v)=2(r+1)-(i+j) < r+1=\dmin$. This concludes the proof. \qed

\subsection{$LPG \setminus mLPG \neq \emptyset$ and $mLPG \setminus LPG \neq \emptyset$}\label{subsec:proper}

Here we show that neither of the classes LPG and mLPG is contained in the other one by providing, for each of these classes a particular graph which is proper to it. 

\begin{Theorem} \label{theo:splitmatching}
Let $G$ be a split matching graph. Then $G \not \in mLPG$, $G \in LPG$ and in this case it is polynomial to find the tree $T$ and the value $\dmax$ associated to $G$.
\end{Theorem}

The proof will follow immediately by the next two lemmas.

\begin{figure}[!ht]
  \begin{center}
 \subfigure[]{\label{fig:matching}	\includegraphics[scale=0.35]{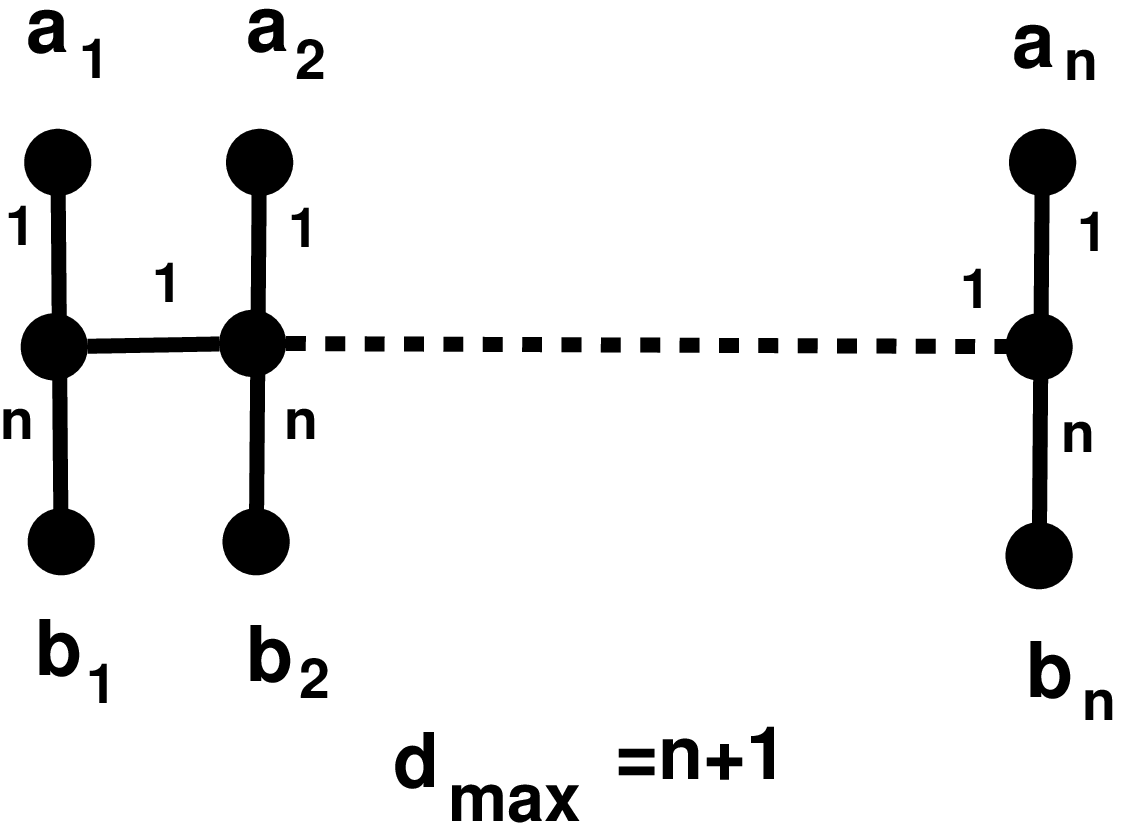}}
 \subfigure[]{\label{fig:matchingbinary}	\includegraphics[scale=0.35]{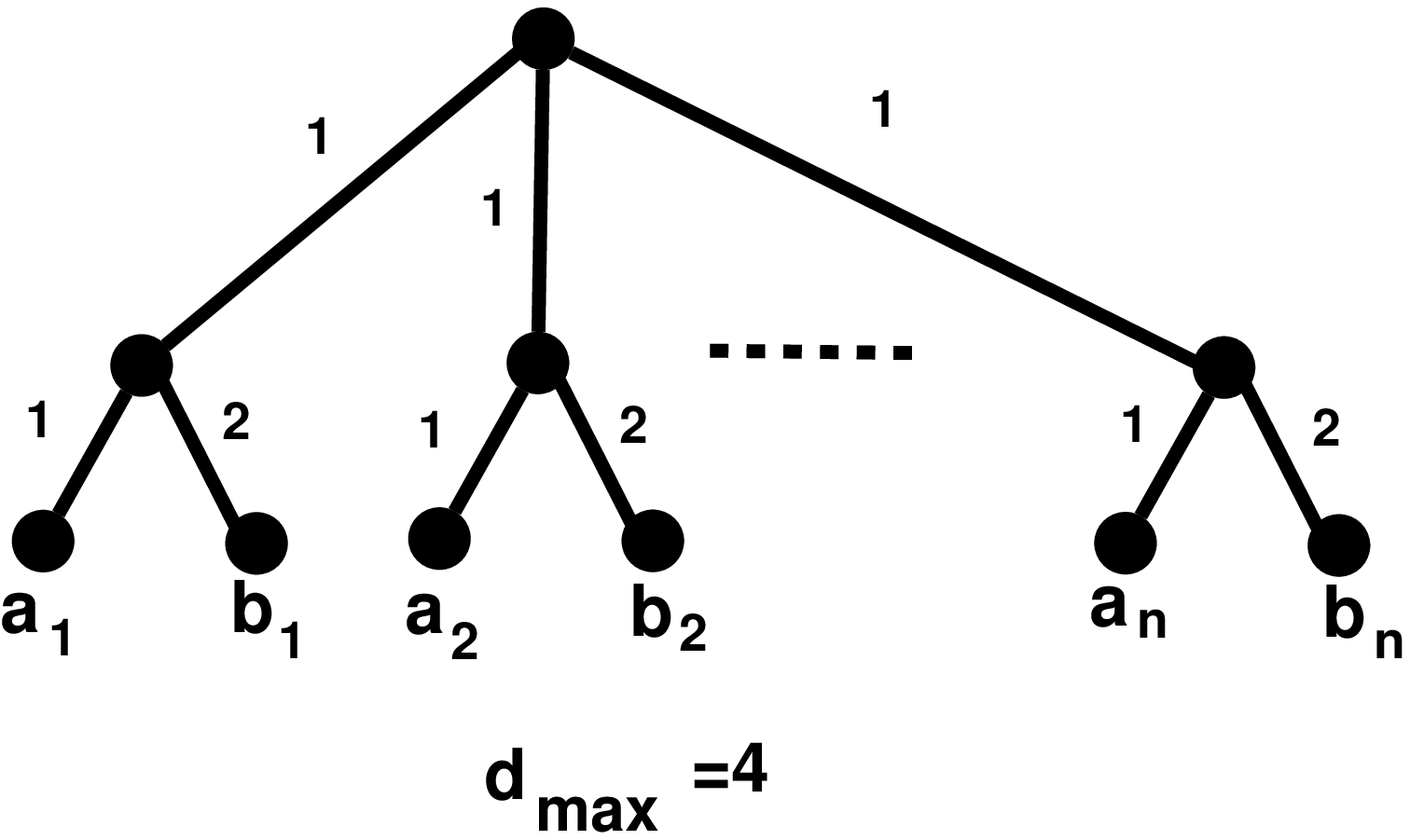}}
  \end{center}
\caption{\footnotesize{(a) A pairwise compatibility caterpillar tree for a split matching graph. (b) A pairwise compatibility tree for a split matching graph.}}
\label{fig:mat}
\end{figure}

\begin{Lemma}\label{lem:matchingtree}
Let $G$ be a split matching graph. Then $G \in LPG$ and it is polynomial to find the tree $T$ and the value $\dmax$ associated to $G$.
\end{Lemma}
\proof
Given a split matching graph $G=(K,S,E)$ with $|K|=|S|=n$, we associate a caterpillar tree $T$ as in Figure \ref{fig:matching}. The leaves $a_i$, corresponding to the vertices $k_i$ of $K$, are connected to the spine with edges of weight $1$ and the leaves $b_i$, corresponding to vertices $s_i \in S$, with edges of weight $n$. It is clear that $G=LPG(T,n+1)$. Indeed, for any two $a_i, a_j$ it holds that $3\leq d_T(a_i,a_j)\leq n+1$, for any two $b_i, b_j$ we have $d_T(b_i,b_j)\geq 2n+1$ and for any $a_i, b_i$ we have $d_T(a_i,b_i)=n+1$ (hence the edge $(k_i,s_i) \in E$) and for any $a_i, b_j$ with $i\neq j$ we have $d_T(a_i,b_j) \geq n+2$ (hence the edge $(k_i,s_j) \not\in E$).\qed
Note that this representation is not unique. Indeed, one can easily check that the binary tree $T$ in Figure \ref{fig:matchingbinary} also is a pairwise compatibility tree of a split matching graph when $\dmax=4$. 

\begin{Lemma}\label{lem:matching2}
Let $G$ be a split matching graph. Then $G \not \in mLPG$.
\end{Lemma}
\proof
Given a split matching graph $G=(K,S,E)$ with $|K|=|S|=n$, we assume by contradiction $G=mLPG(T,\dmin)$. Then let $a_1,a_2,a_3$ be three leaves of $T$ corresponding to three vertices of $K$, $k_1,k_2,k_3$. Without loss of generality let $P_{a_1 a_2}$ be the largest path in the subtree $T_{a_1 a_2 a_3}$. Consider the vertex $s_3$ in $S$ associated to the leaf $b_3$ in $T$, with $(k_3, s_3) \in E$. From Lemma \ref{lem:technical} we deduce that either $d_T(b_3,a_3) \leq d_T(b_3,a_2)$ or $d_T(b_3,a_3) \leq d_T(b_3,a_1)$. The existence of the edge $(k_3,s_3)$ in $G$ implies $d_T(b_3,a_3) \geq \dmin$, therefore one from $(k_1, s_3)$ and $(k_2, s_3)$ must be an edge in $G$, a contradiction. \qed

Analogously, we can show that the set $mLPG \setminus LPG$ is not empty.

\begin{Theorem} \label{theo:splitantimatching}
Let $G$ be a split antimatching graph. Then $G \not \in LPG$,  $G \in mLPG$ and in this case it is polynomial to find the tree $T$ and the value $\dmin$ associated to $G$.
\end{Theorem}

\begin{figure}[!ht]
  \begin{center}
 \subfigure[]{\label{fig:antimatching}	\includegraphics[scale=0.35]{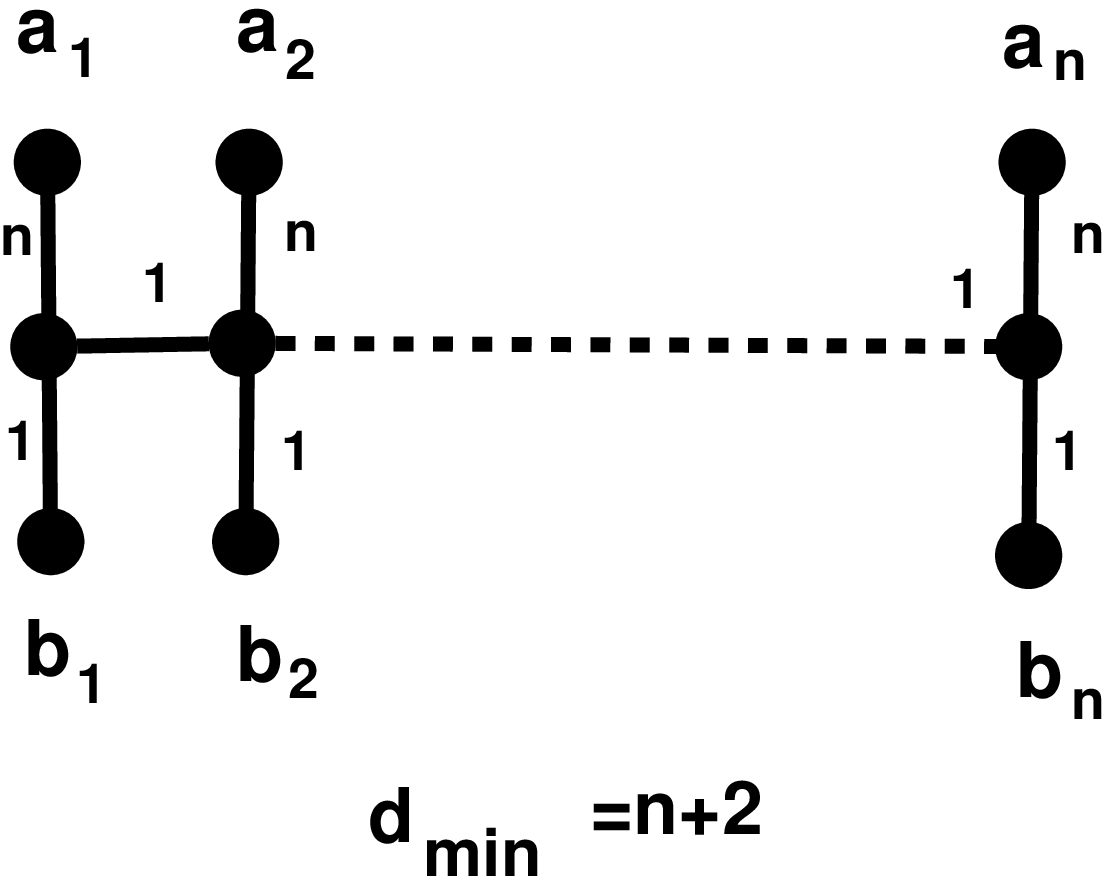}}
 \subfigure[]{\label{fig:antimatchingbinary}	\includegraphics[scale=0.35]{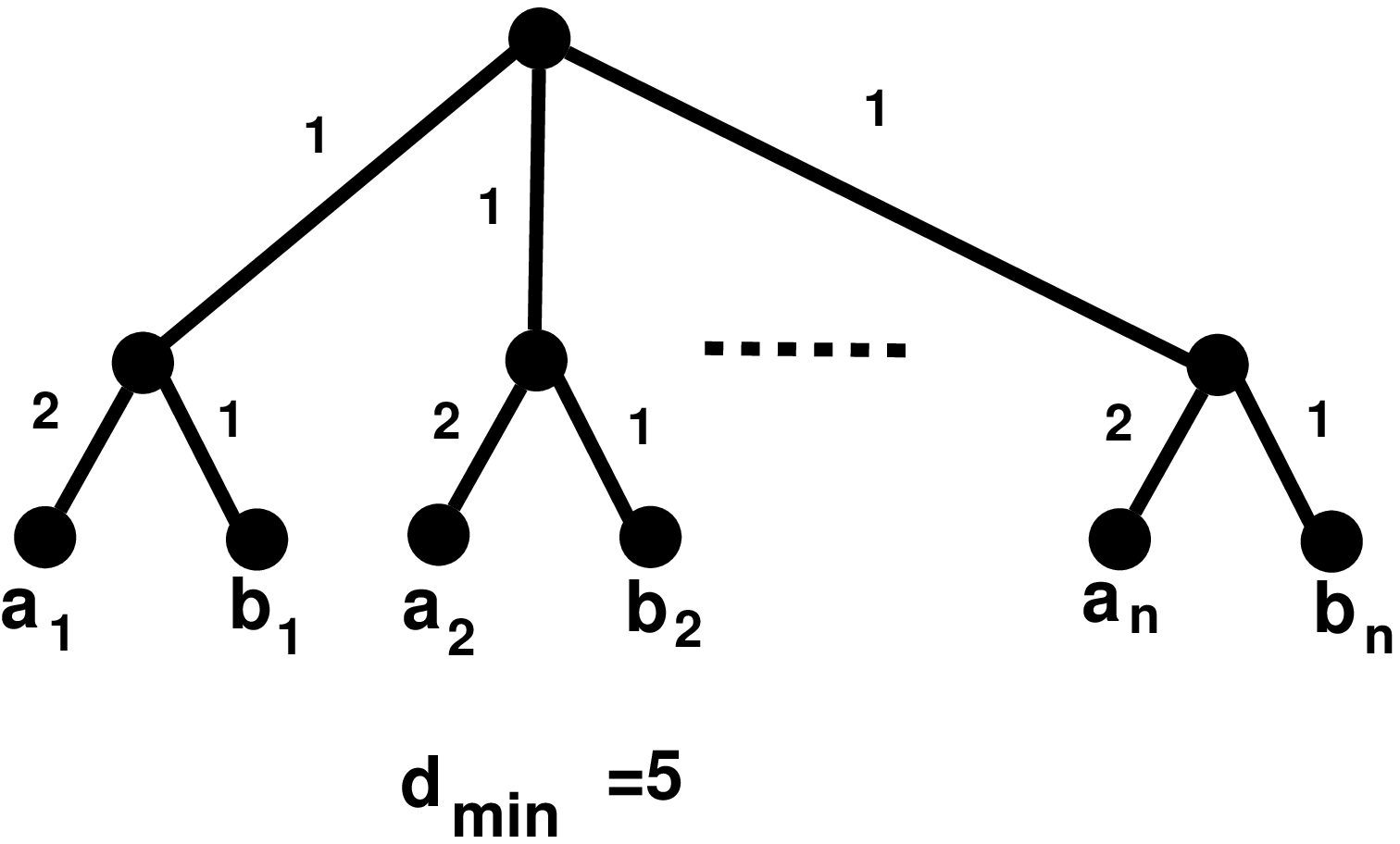}}
  \end{center}
\caption{\footnotesize{(a) A pairwise compatibility caterpillar tree for a split antimatching graph. (b) A pairwise compatibility tree for a split antimatching graph.}}
\label{fig:anti}
\end{figure}

For the sake of brevity we omit the proof of this theorem, that follows using arguments similar to those in the proofs of Lemmas \ref{lem:matchingtree} and \ref{lem:matching2}. The tree $T$ associated to a split antimatching graph is one from the ones depicted in Figure \ref{fig:anti}.

\section{Split Matrogenic Graphs}\label{sec:matrogenic}

In Section \ref{sec:relations}, when studying the relations among the three classes PCG, LPG and mLPG, we have dealt with threshold graphs, split matchings and split antimatchings. All these graphs are split matrogenic graphs (cfr. definition later). For this reason, it is natural to ask whether split matrogenic graphs are PCG or not. This section is devoted to answer this question.

\begin{Definition}
A {\em split matrogenic} graph is the composition of $t$ split graphs $G_i=(K_i,S_i,E_i)$ with $i=1, \ldots, t$ such that: either $G_i$ is a split matching or $G_i$ is a split antimatching or $K_i=\emptyset$ (and $G_i$ is called {\em stable graph}) or $S_i=\emptyset$ (and $G_i$ is called {\em clique graph}).
\end{Definition}

In order to make easier the exposition, we introduce two subclasses of split matrogenic graphs.

\begin{Definition}
Given a sequence of $t$ split graphs $G_i=(K_i,S_i,E_i)$ with $i=1, \ldots, t$, we say the graph $H= G_1 \circ \ldots \circ G_t$ is a {\em split matching (antimatching) sequence} if each of the graphs $G_i$ is either a split matching (antimatching), or a  stable graph or a clique graph. 
\end{Definition}

\begin{figure}[!ht]
  \begin{center}
 \subfigure[]{\label{fig:matchingbinaryEmpty} \includegraphics[scale=0.4]{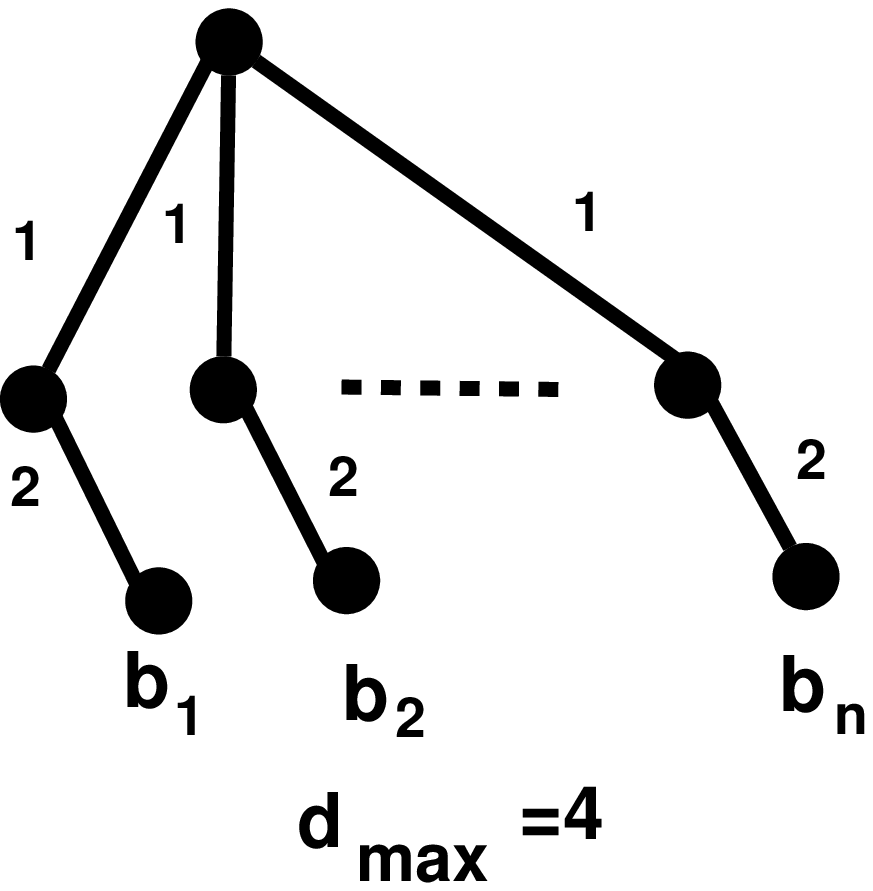}}
 \subfigure[]{\label{fig:antimatchingbinaryEmpty}\includegraphics[scale=0.4]{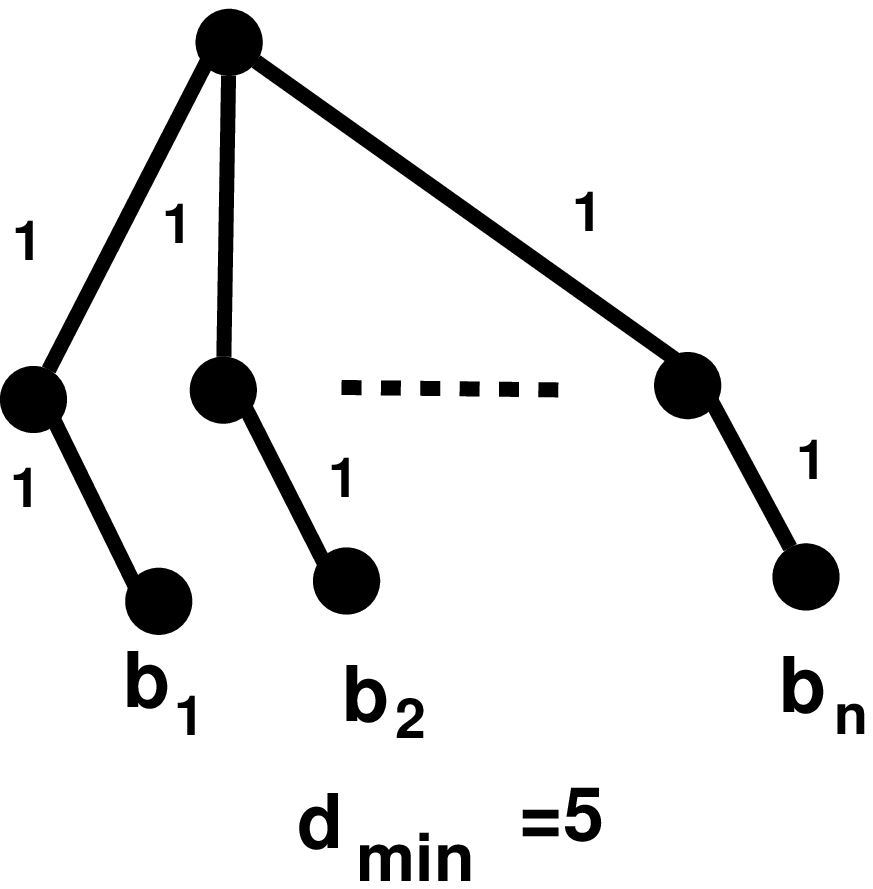}}
  \end{center}
\caption{\footnotesize{ The pairwise compatibility tree for a stable graph $G$ with $n$ vertices when it is considered as (a) a split matching graph  (b) a split antimatching graph. }}
\label{fig:matchingAntimatchingEmpty}
\end{figure}

We first prove that split matching sequences and split antimatching sequences are PCG. In both these proofs, in the construction of the pairwise compatibility tree, we will make use of the constructions depicted in Figure \ref{fig:matchingbinary} and Figure  \ref{fig:antimatchingbinary}, respectively. Finally, we want to point out that a clique graph (a stable graph) can be considered both as a split matching or a split antimatching graph and in each case the pairwise compatibility tree is constructed in the same way, where only leaves $a_i$ (respectively $b_i$) appear. In Figure \ref{fig:matchingAntimatchingEmpty} the pairwise compatibility tree is given for a stable graph $G$ when it is considered first as a split matching graph and next as a split antimatching graph.

\begin{figure}[!ht]
  \begin{center}
 \subfigure[]{\label{fig:matchingsequenceBi} \includegraphics[scale=0.4]{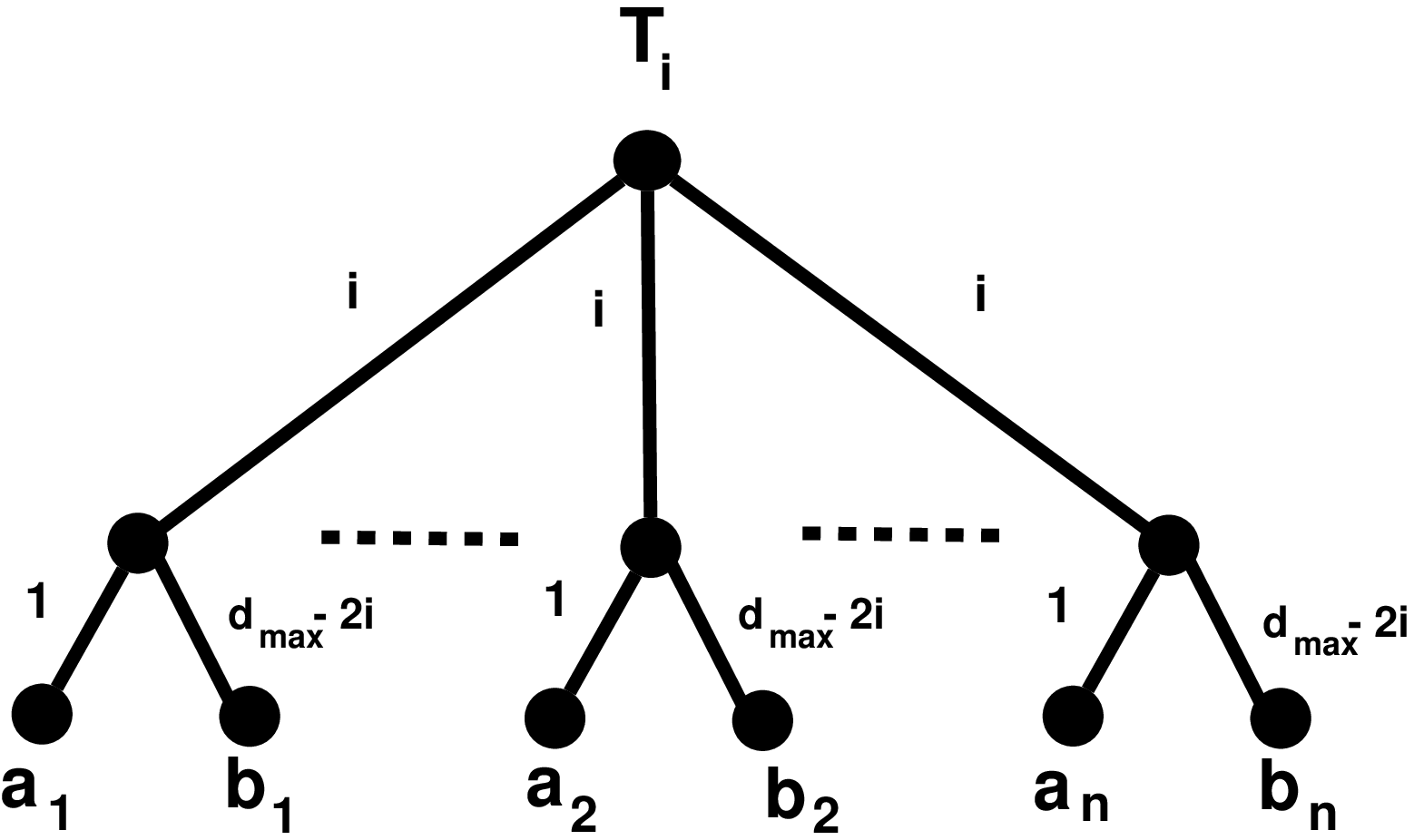}}
 \subfigure[]{\label{fig:matchingsequenceTree}\includegraphics[scale=0.4]{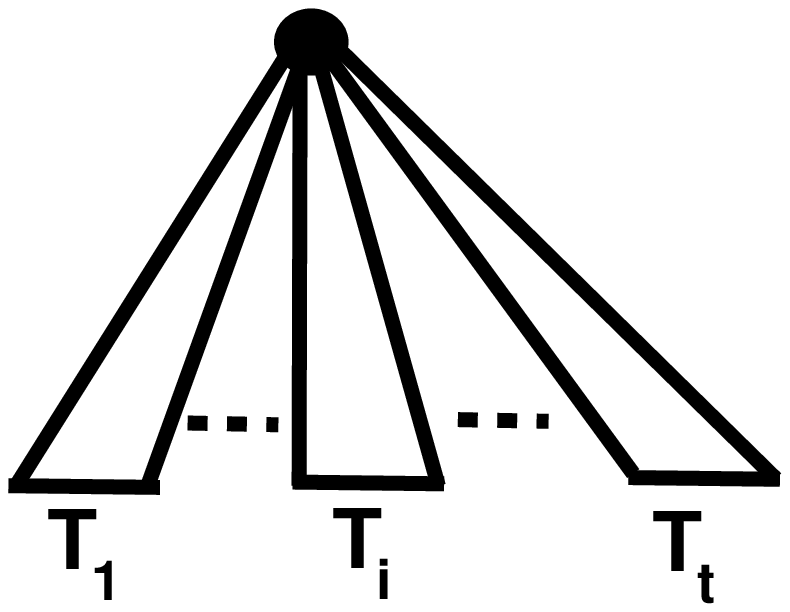}}
  \end{center}
\caption{\footnotesize{(a) The pairwise compatibility tree for the split matching graph $G_i$.  (b) The pairwise compatibility tree for the split matching sequence $H$. }}
\label{fig:matchingsequence}
\end{figure}

\begin{Theorem}\label{theo:matchingSequence}
Let $H$ be a split matching sequence. Then $H \in LPG$ and it is polynomial to find the tree and the value $\dmax$ associated to $H$.
\end{Theorem}
\proof
Let $H= G_1 \circ \ldots \circ G_t$ be a  split matching sequence. For each graph $G_i$ we define a tree $T_i$ as shown in Figure \ref{fig:matchingsequenceBi} (where the leaves $a_i$ ($b_i$) may not possibly appear if $G_i$ is a stable (clique) graph). It is clear that $G_i=LPG(T_i, \dmax)$ where $\dmax$ is a value to be defined later, but surely greater than or equal to $2(i+1)$. Indeed, let $a_1, \ldots a_n$ be the leaves of $T_i$ corresponding to vertices of $K_i$ and $b_1, \ldots, b_n$ those corresponding to vertices of $S_i$. For any two leaves $a_r, a_s$ it holds that $d_{T_i}(a_r,a_s)=2+2i\leq \dmax$ and for any two $b_s, b_r$ we have $d_{T_i}(b_r,b_s)=2 \dmax-2i \geq \dmax +2i+2-2i > \dmax$. Finally, for any two leaves $a_s, b_s$ that correspond to an edge of the matching their distance is $\dmax-2i+1 \leq \dmax$ and for any two leaves corresponding to a non edge $a_r, b_s$ their distance is $\dmax +1$. 

In order to prove that $H \in LPG$,  we define a new tree $T$ starting from the trees $T_1, \ldots, T_t$, simply by contracting all their roots to a single vertex as shown in Figure \ref{fig:matchingsequenceTree}. We claim that $H=LPG(T, \dmax)$ where we set $\dmax = 2(t+1)$. In order to prove it, consider two graphs $G_i$ and $G_j$ with $i < j$. Let  $a, a', b$ and $b'$ be four distinct leaves  corresponding to vertices in $K_i, K_j, S_i$ and $S_j$ respectively. 
Observe that the vertices in $K_i$ are connected to all the other vertices in $K_j \cup S_j$ as the distances in $T$ are $d_T(a,a') = 1+ i+j+1 \leq 2(j+1) \leq \dmax$ and $d_T(a,b') = 1+i+j+\dmax-2j= \dmax +(i-j+1) \leq \dmax$ (as $j \geq i+1$). Finally, any vertex in $S_i$ is not connected to any vertex $K_j$ and to any vertex $S_j$ as in these cases the distances are $d_T(b,a') = \dmax-2i+i+j+1>\dmax$ (as $j \geq i+1$) and $d_T(b,b') =\dmax-2i+i+j+\dmax-2j \geq 2\dmax-2j >\dmax$. \qed

\noindent
Using similar arguments we prove the following result.

\begin{Theorem}\label{theo:antimatchingSequence}
Let $H$ be a split antimatching sequence. Then  $H \in mPCG$ and it is polynomial to find the tree and the value $\dmin$ associated to $H$.
\end{Theorem}
\proof

\begin{figure}[!ht]
  \begin{center}
\includegraphics[scale=0.4]{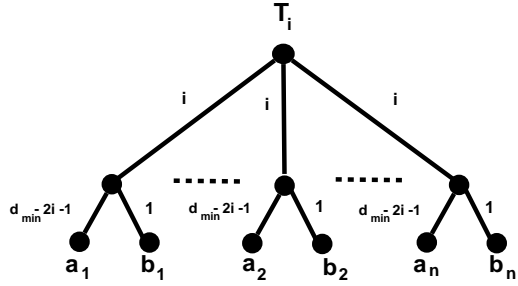}
  \end{center}
\caption{\footnotesize{The pairwise compatibility tree for the split antimatching graph $G_i$. }}
\label{fig:antimatchingsequence}
\end{figure}

The proof follows the same lines of the proof of Theorem \ref{theo:matchingSequence}. Let $H= G_1 \circ \ldots \circ G_t$ be a  split antimatching sequence. We will associate to each split antimatching graph $G_i$ a tree $T_i$ as depicted in Figure \ref{fig:antimatchingsequence}. 
We prove that $G_i=mLPG(T_i, \dmin)$ where $\dmin$ is a value to be defined later, but surely greater than or equal to $2(i+1)+1$. Indeed, let $a_1, \ldots a_n$ be the leaves of $T_i$ corresponding to vertices of $K_i$ and $b_1, \ldots, b_n$ those corresponding to vertices of $S_i$. For any two leaves $a_r, a_s$ it holds that $d_{T_i}(a_r,a_s)=2 \dmin -2i-2= \dmin + (\dim -2(i+1)) \geq \dmin$ and for any two $b_s, b_r$ we have $d_{T_i}(b_r,b_s)=2i+2 < \dmin$. Finally, for any two leaves $a_s, b_s$ that correspond to an edge of the antimatching their distance is $\dmin-2i-1+2i+1=\dmin$ and for any two leaves corresponding to a non edge $a_r, b_r$ their distance is $\dmin-2i$. 

We define the tree $T$ starting from the trees $T_1, \ldots, T_t$, in the same way we have done in the previous theorem (see Figure \ref{fig:matchingsequenceTree}). Using the same arguments it is not difficult to check that $H=mLPG(T, \dmin)$ where we have set $\dmin = 2(t+1)+1$. 
Indeed, consider two graphs $G_i$ and $G_j$ with $i < j$. Let  $a, a', b$ and $b'$ be four distinct leaves  corresponding to vertices in $K_i, K_j, S_i$ and $S_j$ respectively. 
Observe that the vertices in $K_i$ are connected to all the other vertices in $K_j \cup S_j$ as the distances in $T$ are 
$d_T(a,a') = \dmin-i-1+\dmin-j-1= \dim + (\dmin- (i+j+2)) \geq \dmin$ and $d_T(a,b') = \dmin-i-1+j+1\geq \dmin$. Finally, any vertex in $S_i$ is not connected to any vertex $K_j$ and to any vertex $S_j$ as in these cases the distances are $d_T(b,a') = 1+i+j+\dmin-2j-1=\dmin+i-j < \dmin$ (as $j \geq i+1$) and $d_T(b,b') =1+i+j+1 < 2(j+1) \leq 2(t+1) < \dmin$. \qed

\begin{figure}[!ht]
  \begin{center}
\includegraphics[scale=0.34]{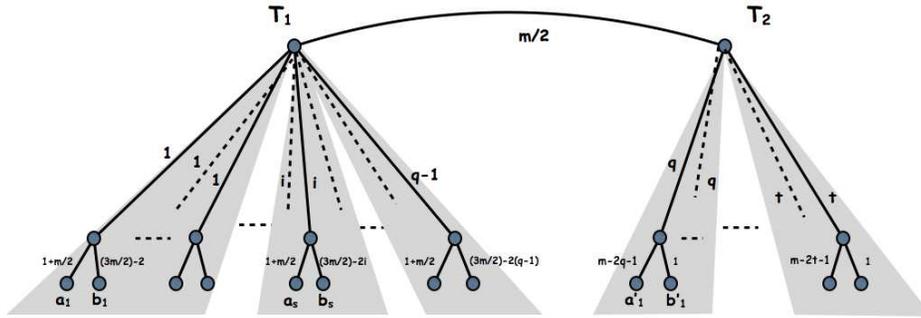}
  \end{center}
\caption{\footnotesize{The pairwise compatibility tree for the split matrogenic graph $H$. }}
\label{fig:matchingAntimatching}
\end{figure}

\begin{Theorem}
Let $H=G_1 \circ \ldots \circ G_t$ be a split matrogenic graph such that for any split matching graph $G_i$ and for any split antimatching graph $G_j$ it holds  that $i<j$. Then $H \in PCG$ and it is polynomial to find the tree and the values $\dmin, \dmax$ associated to $H$.
\end{Theorem}
\proof
Let $H=G_1 \circ \ldots \circ G_t$. It is clear that if none of the graphs $G_i$ is a split matching (a split antimatching) the proof trivially follows from Theorem \ref{theo:matchingSequence} (Theorem \ref{theo:antimatchingSequence}). Hence, let $G_q$, $1 < q \leq t$, be the first occurrence of a split matching graph. Then, the graphs $H_1= G_1 \circ \ldots \circ G_{q-1}$ and $H_2= G_q \circ \ldots \circ G_t$ are a split matching sequence and a split antimatching sequence, respectively.
Then, let $H_1=LPG(T_1,M)$ where the tree is constructed in the same way as in the proof of Theorem \ref{theo:matchingSequence} and $M=2(t+1)+1$ (recall that in the proof of Theorem \ref{theo:matchingSequence}  we only need $M$ to be a value greater that $2q$). Similarly, according to the Theorem  \ref{theo:antimatchingSequence}, $H_2=mLPG(T_2,m)$ and $m=2(t+1)+1$ (note that we choose to have $m=M$)). We modify $T_2$ in such a way that of the weights of the edges outcoming form the root start from value $q$ instead of from $1$ and the other edges are modified accordingly. This is not restrictive, as $T_2$ results as if $H_2$ was the composition of $t$ split antimatching graphs whose the first $q-1$ are empty graphs.


We construct the pairwise compatibility tree $T$ by joining the roots of $T_1$ and $T_2$ with an edge of weight $m/2$. We set $\dmin=m$ and $\dmax=2m$. We modify the weights of the resulting tree increasing by $m/2$ the weight of any edge incident to a leaf in $T_1$.  Observe that in this way the distance of any two leaves in $T_1$ is increased by $m$. This means that two leaves correspond to vertices of an edge in $H_1$ if and only if their distance is less than or equal to $M+m=2m$. Furthermore, the maximum distance of any two leaves in $T_2$ is less than or equal to $2m-2t < 2m$ meaning that they correspond to vertices of an edge in $H_2$ if and only if their distance is greater than or equal to $m$. 

We claim that $H=PCG(T,2m,m)$ (recall that $m=2(t+1)+1$). We have already shown that the pairwise compatibility constraints hold for any two leaves that correspond to two vertices of the same graph $H_1$ or  $H_2$. It remains to show it also holds for two leaves  where one corresponds to a vertex in $H_1$ and the other to $H_2$. To this purpose, let $a_i$ and $b_i$ be two distinct leaves in $T_1$, connected to the root with edges of weight $i$ and  corresponding to vertices of the clique and the stable graph of $H_1$, respectively. Similarly let $a'_j, b'_j$ be two distinct leaves in $T_2$, connected to the root with edges of weight $j$ and corresponding to vertices in the clique and the stable graph of $H_2$, respectively. The following hold:

\begin{itemize}
\item[a)] $d_T(a_i,a'_j)=2m+i-j$ and as $i < j$ and $m>j$ then $m \leq 2m+1+i-j\leq 2m$. Hence, the corresponding vertices of $a_i, a'_j$ in $H$ are connected.
\item[b)] $d_T(a_i,b'_j)=m+1+i+j+1$ and as $m=2t+3 \geq i+j+2$ then $m \leq m+i+j+2 \leq 2m$.  Hence, the corresponding vertices of $a_i, b'_j$ in $H$ are connected.
\item[c)] $d_T(b_i,a'_j)=2m-i+m-j-1$ and as $m=2t+3 \geq i+j+2$ then $2m+(m-i-j-1) > 2m$. Hence, the corresponding vertices of $b_i, a'_j$ in $H$ are not connected.
\item[d)] $d_T(b_i,b'_j)=2m-i+j+1$ and as $i<j$ then $2m+(i-j+1) > 2m$. Hence, the corresponding vertices of $b_i, b'_j$ in $H$ are not connected.
\end{itemize}

This, concludes the proof. \qed

It seems that the order of appearance of a matching or an antimatching sequence in a split matrogenic graph is somehow strictly related to the pairwise compatibility property.  Indeed, in spite our efforts, the following problem remains open.

\textit{Problem:} Let $H=G_1 \circ \ldots \circ G_t$ be a split matrogenic graph such that for any split antimatching graph $G_i$ and for any split matching graph $G_j$ it holds  that $i<j$. Is $H$ a PCG ?

If this problem has an affermative answer then it should not be difficult to prove that all split matrogenic graphs are PCG. Otherwise, the separation between the split matrogenic graphs that are PCG and those that are not, would be perfectly known.

\section{Conclusions and Open Problems}\label{sec:conclusion}

In this paper we analyze the class of PCG with particular attention to two particular subclasses resulting when the pairwise compatibility constriants are relaxed. Hence, we consider the sublasses LPG and mLPG, resulting from the the cases where $\dmin=0$ and $\dmax=+\infty$, respectively. We study the relations between the classes PCG, LPG and mLPG. In particular, we show that the union of LPG and mLPG does not coincide with the whole class PCG, their intersection is not empty, and that neither of the classes LPG and mLPG is contained in the other.  The graphs considered in these considerations are particular cases of the more general class of split matrogenic graphs. Hence,  we attempt to determine whether the class of split  matrogenic graphs belongs to PCG class. We prove that many split matrogenic graphs are PCG. However, the membership to PCG class of one particular split matrogenic graph remains an open problem.  

It should be stressed that up to date, the pairwise compatibility has been investigated only for a few classes of graphs, thus determine this property for many graph classes remains an open problem.  It is worth to mention that in \cite{Kal03} is shown that the clique problem can be solved in polynomial time for the class of compatibility graphs if we are able to construct in polynomial time a weighed tree that generates the pairwise compatibility graph.  In view of this, it seems even more interesting to completely identify the Pairwise Compatibility graphs class.


\end{document}